*atmosphere*

Article

# Could macroscopic dark matter (macros) give rise to mini-lightning flashes out of a blue sky without clouds?

Vernon Cooray [1,*], Gerald Cooray [2], Marcos Rubinstein [3] and Farhad Rachidi [4]

1   Department of Electrical Engineering, Uppsala University, 752 37 Uppsala, Sweden.
2   Karolinska Institute, Stockholm, 171 77 Solna, Sweden.
3   HEIG-VD, University of Applied Sciences and Arts Western Switzerland, 1401 Yverdon-les-Bains, Switzerland
4   Electromagnetic Compatibility Laboratory, Swiss Federal Institute of Technology (EPFL), 1015 Lausanne, Switzerland

**Abstract:** A recent study pointed out that macroscopic dark matter (macros) traversing through the earth's atmosphere can give rise to hot and ionized channels similar to those associated with lightning leaders. The authors of the study investigated the possibility that when such channels created by macros pass through a thundercloud, lightning leaders may be locked in by these ionized channels creating lightning discharges with perfectly straight channels. They suggested the possibility of detecting such channels as a means of detecting the passage of macros through the atmosphere. In this paper, we show that macros crossing the atmosphere under fair weather conditions could also give rise to mini-lightning flashes with current amplitudes in the order of few hundreds of Amperes. These mini-lightning flashes would generate a thunder signature similar to or stronger than those of long laboratory sparks and they could also be detected by optical means. As in the case of thunderstorm-assisted macro lightning, these mini-lightning flashes are also associated with straight channels. Moreover, since the frequency of mini-lightning flashes is about thirty times greater than the macro-generated lightning flashes assisted by thunderstorms, they could be used as a means to look for the paths of macroscopic dark matter crossing the atmosphere.

## 1. Introduction

In a recent paper, Starkman et al. [1] have explored and discussed various ways that the passage of macroscopic dark matter (macros) across the Earth could be detected. Since macros interact strongly with matter, they pointed out that the passage of them through the atmosphere can lead to the ionization of air creating an ionized trail along their trajectory. Utilizing this fact, Starkman et al. [1] came up with a novel idea to detect the presence and propagation of macros through the atmosphere. The idea they have put forward is the following. The creation of a lightning flash between the cloud and the ground is preceded by the formation of a conducting channel between them. This is mediated by a discharge called leader that travels generally downward generating a conducting channel between the cloud and the ground. Charges will be induced on the leader channel to maintain its potential close to that of the cloud. Since the channel of the leader discharge is at a higher potential with respect to the ground, when it reaches the ground an ionization wave that carries the ground potential travels along it from ground to cloud. This is called the return stroke.   Starkman et al. [1] pointed out that a macro passing through the atmosphere will give rise to an ionized channel with temperature and electron densities similar to those of a lightning leader. They hypothesized that when a macro passes through a thundercloud, a lightning leader could be locked in by this ionized channel and travel towards the ground along the ionized channel giving rise to a lightning flash. Since the ionized channel is created by the macro and not by the stochastic processes associated with the leader discharge, an important and distinguishable characteristic of this lightning channel would be that it is completely straight. This is indeed the case if the speed of the leader discharge is slower than the speed of the macros. Otherwise, the leader discharge will break away from the channel of the macros creating a tortuous channel similar to that of a normal lightning. Commenting on the hypothesis presented in this paper, Cooray et al. [2] pointed out that there is no need for a lightning leader to be locked in by this macros channel but the creation of a conducting channel in the electric field of the cloud itself, due to its polarization in the background electric field, will transform the channel of the macro to a leader channel with properties more or less similar to those of a normal lightning leader.

Since the creation of a lightning flash by a macro requires its path to cross a thunderstorm, the number of lightning flashes created by macros will depend on the number of thunderstorms taking place at any time on the surface of the Earth. The typical size of a thundercloud is about 20 km in diameter and about 2000 thunderstorms are active at any given time. Based on this data, Starkman et al. [1] estimated that the macro induced lightning flash rate will be about $10^{-6}$/s. This is much smaller than the rate normal lightning flashes striking the earth which is about 100/s. Thus, macro induced lightning channels, if present, will be hidden inside the flux of normal lightning flashes taking place in the Earth's atmosphere. Moreover, depending on the speed of the macro, the large concentration of electric charge at the tip of the macro channel induced by the thundercloud electric field may cause the leader discharge to escape from the macro channel, thus camouflaging the macro lightning channel as a normal tortuous lightning channel.

Now, since the ionization of the macro channel is not related to the presence of a thundercloud, the creation of an ionized channel inside the earth's atmosphere by a macro takes place irrespective of whether the macro is passing through a thunderstorm region or through a fair-weather region. The goal of this paper is to show that, even in the case of a macro propagating across the atmosphere under fair weather conditions, the macro could give rise to a mini-lightning flash which could be observed at close range and be detected optically, especially during nighttime.

## 2. Hypothesis

The primary reason for the creation of a lightning flash when a macro is passing through the Earth's atmosphere in a region covered by an active thundercloud is the polarization of the macro channel by the large electric field that exists along the path of the macro. This electric field, which is in the order of a few tens of kV/m, will induce charges on the conducting macro channel, creating a situation identical to that of a propagating leader discharge. Indeed, the creation of a normal leader also takes place in a similar manner. The electrical processes taking place at the tip of the leader channel generate a conducting channel of several tens of meters of length ahead of the leader channel. The newly created channel section will be charged by the action of the local electric field. This charging process manifests as a bi-directional current pulse that propagates forward along the newly created conducting channel and backwards along the existing leader channel [3]. The current propagating along the already existing leader channel (i.e., the backward-propagating current) helps to maintain its conductivity so that a potential comparable to the cloud potential is maintained at its head. Thus, the main function of the leader discharge is the creation of a conducting channel in air and the background electric field will redistribute the electrical charges so that a high potential is maintained at the leader head. During this process, charges of one polarity remain on the leader channel while charges of the opposite polarity travel along the leader channel to its origin inside the cloud. The same process will take place whenever there is a background electric field present in the region where the conducting channel is formed. However, the magnitude of the accumulated charge and the resulting peak of the return stroke current will be decided by the strength of the background electric field.

Interestingly, thunderstorms and lightning not only change the electric field in their vicinity but they mediate in creating a global electric field that is present around the globe even under fair weather conditions [4, 5]. This fair-weather electric field is directed towards the ground indicating the presence of negative charge on the earth. It is assumed that this negative charge is maintained by the action of lightning flashes and corona discharges taking place under thunderclouds. This electric field is about 100 V/m close to ground and it decreases nearly exponentially with height since the conductivity of the atmosphere increases with height. At the height of the ionosphere (i.e., 60 – 90 km) the conductivity becomes so high the electric field goes down to negligible values. This fair-weather electric field makes the earth's ionosphere be at a higher potential with respect to the ground. It is estimated that this ionospheric potential is about 250 - 400 kV [5, 6]. Most of the contribution to this potential is coming from the fair-weather electric field located below about 20 km height.

Now, consider a macro passing through the Earth's atmosphere creating an ionized channel, with a conductivity comparable to that of a normal leader channel, between the ionosphere and the Earth's surface. As the macro approaches the ground, the conducting channel will be polarized due to the fair-weather electric field, generating currents along this conducting channel leading to the deposition of charge along the channel. This charge will be of positive polarity since the electric field is directed downwards towards the ground (i.e., the earth is negatively charged with respect to the ionosphere). Of course, the magnitude of this charge would be much less than that of a macro channel passing through a thundercloud, but, as we will show in the next section, the charging of the macro channel by the fair-weather electric field can give rise to a mini-lightning flash whose effects could be observable. We call it a mini-lightning flash since its current is significantly (about 2 orders of magnitude) smaller than that of a normal lightning flash.

## 3. Charging of a macro channel in the fair weather electric field

According to [5], the fair-weather electric field as a function of height can be represented by the equation

$$E(z) = 93.8e^{-4.527z} + 44.4e^{-0.375z} + 11.8e^{-0.121z} \tag{1}$$

In the above equation, $z$ is the height above the surface of the earth in km and $E(z)$ is the electric field at height $z$. This expression represents the fair-weather electric field in the mid latitudes for heights below about 60 km. The electric field predicted by (1) and the atmospheric potential resulting from this electric field change with height are shown in Figure 1. Note that the electric field decreases rapidly with height, due to the increasing conductivity of the atmosphere, and this electric field raises the upper atmosphere to a potential of about 250 kV. Observe also that nearly the full atmospheric potential is reached within about 20 km from ground level due to the very small contribution to the atmospheric potential by the electric field at heights larger than about 20 km. In the analysis presented here, we will continue to use Equation (1) to represent the fair-weather electric field.

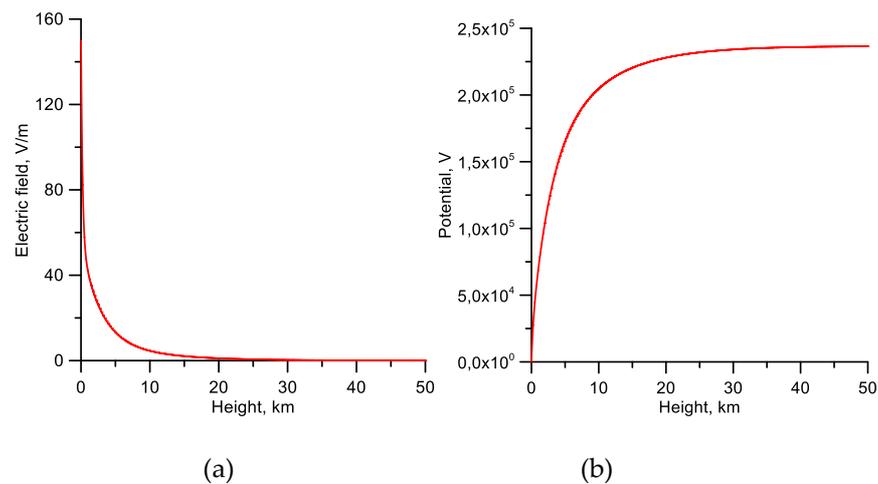

**Figure 1.** (a) The variation of the fair-weather atmospheric electric field with height. (b) The variation of atmospheric potential with height.

In the analysis, the macro channel radius is taken to be 10 cm, comparable to that of a normal lightning leader channel. This channel radius is assumed to be the region where the induced charge is located. It is important to point out that the radius inside which the charge is located is controlled by the electric field at the surface of the charged macro channel. If this electric field is larger than about $3\times10^6$ V/m, the breakdown electric field at atmospheric pressure, the charge will leak out of the channel up to a radial length where the electric field is reduced to about $5\times10^5$ V/m. This is the threshold electric field for streamer propagation [7]. As one can see later, the charge per unit length that would be induced on a 10-cm radius channel by the fair-weather electric field is about a few micro Coulombs. Thus, this channel radius is capable of accommodating this charge without leakage. Figure 2 depicts the distribution of the charge per unit length accumulated along the channel just before the macro channel reaches the earth's surface (i.e., within 2 m from the earth's surface). This accumulation of positive charge along the macro channel will generate a current that propagates up along the macro channel transporting negative charge towards the ionosphere. We will make a rough estimation of this channel current too in the next section.

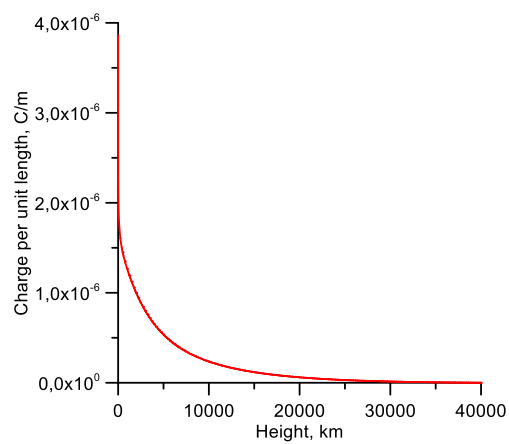

**Figure 2.** The charge per unit length induced on the macro channel when its tip is at ground level (in fact 2 m above ground) by the fair-weather atmospheric electric field.

The magnitude of the charge per unit length induced on the macro channel is controlled by the background electric field. Note that the charge per unit length induced on the macro channel when it is in the vicinity of the earth's surface is about 3 μC/m. When this conducting channel reaches the ground, the neutralization of the accumulated charge, which indeed is at a higher potential with respect to ground, will give rise to an ionization wave similar to that of a return stroke that propagates up along the channel. The next task is to estimate the currents associated with this ionization wave, i.e., the current associated with the mini-lightning flash.

## 4. The currents generated in the macro channel

One question that we have to address here is the relationship between the current generated in the macro channel when it reaches the ground and its relationship to the charge per unit length induced on the channel. Let us consider the dart leader-return stroke channel. Thanks to the triggered lightning experiments, we have more information on the relationship between the charge on the dart leader channel and the peak current of the subsequent return stroke that results when this charge is neutralized [8]. Unfortunately, we do not have a direct experimental procedure to estimate the charge per unit length on the dart leader channel. However, an estimation of this parameter can be made using the close electric fields measured at several distances from the triggered lightning channel. These estimates give a value of about 70 µC/m over the first km or so of the dart leader channel associated with a typical subsequent return stroke peak current of 12 kA [8]. Since the charge on the channel increases as the tip of the dart leader approaches the ground, the charge per unit length close to the ground level could be higher than this average estimation. Theoretical estimations on the relation between the return stroke peak current and the charge per unit length at the bottom of the dart leader channel also suggest that a 12-kA return stroke current peak is associated with about 100 µC/m of charge [7]. This indicates that the macro channel with a charge per unit length of about 3 µC/s close to ground will give rise to a mini-return stroke with a peak current of about 300 A. In order to support this conclusion, in the next paragraph we will illustrate how to get an estimate for the currents generated in the macro channel using the procedures commonly used to estimate the dart leader and return stroke currents using the charge distribution of the dart leader channel.

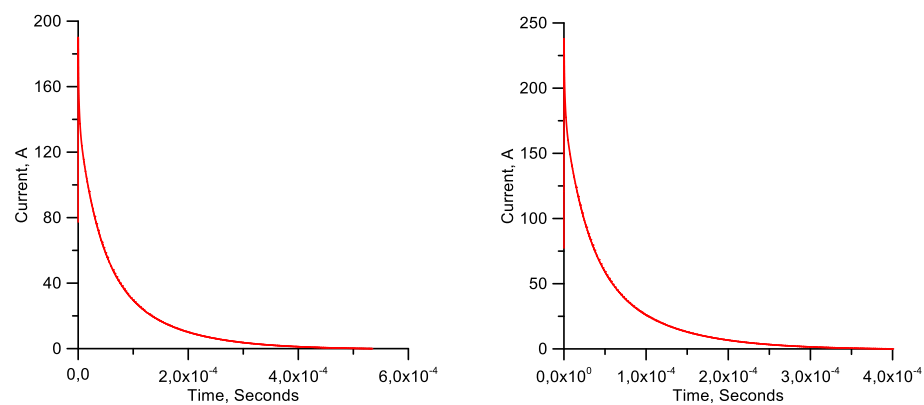

**Figure 3.** (a) The current generated at ground level by the mini-lightning return stroke travelling with a speed $1.0 \times 10^8$ m/s along the macro channel. (b) Same as in (a) except that the speed of propagation is $1.5 \times 10^8$ m/s.

The procedure to estimate the currents resulting from the neutralization of the leader channel is to treat the discharging process as a current-generating mechanism [9]. The discharging process of each channel element will give rise to a current pulse that travels along the leader channel to ground. The return stroke current is a result of the combined action of the discharge currents from the different channel elements. In the analysis, the discharging process is usually assumed to be exponential in nature, resulting in an exponentially decaying discharge current. The amplitude of the discharge current is determined by the charge neutralized in the channel element. A rough estimation of the dis-charge decay time constant can be obtained by using the fact that the risetime of the current at the channel base is approximately equal to the discharge decay time constant. In the case of subsequent return strokes, the risetime of the current is in the order of 0.1 µs and this value can be used in the analysis as the value for the discharge decay time constant. Using this procedure and assuming that the discharging process is similar in both the macro channel and the dart leader channel, we can estimate the current at the base of the macro channel that results during the neutralization process. The resulting current at the base of the macro channel is shown in Figure 3. Note that the estimated current is about 200 A. This is in agreement with the previous estimation. Since the charge on the macro channel decreases with height, the peak current of the mini return stroke also decreases with increasing height. The current waveforms at 5 km and 10 km heights are shown in Figure 4. It is important to point out that the peak of the macro current depends on the speed of the mini return stroke: It increases with increasing speed. Figures 3a and 3b depict the channel-base current for return stroke speeds of $1.0 \times 10^8$ and $1.5 \times 10^8$ m/s, respectively. The total charge brought down by the mini return stroke is about $10^{-2}$ C which is about 50 to 100 times less than that of typical subsequent strokes. Based on these data, we conclude that macro channels generated under fair weather conditions, i.e., out of the blue sky, can generate mini-lightning flashes with amplitudes about 50 times less than typical subsequent return strokes.

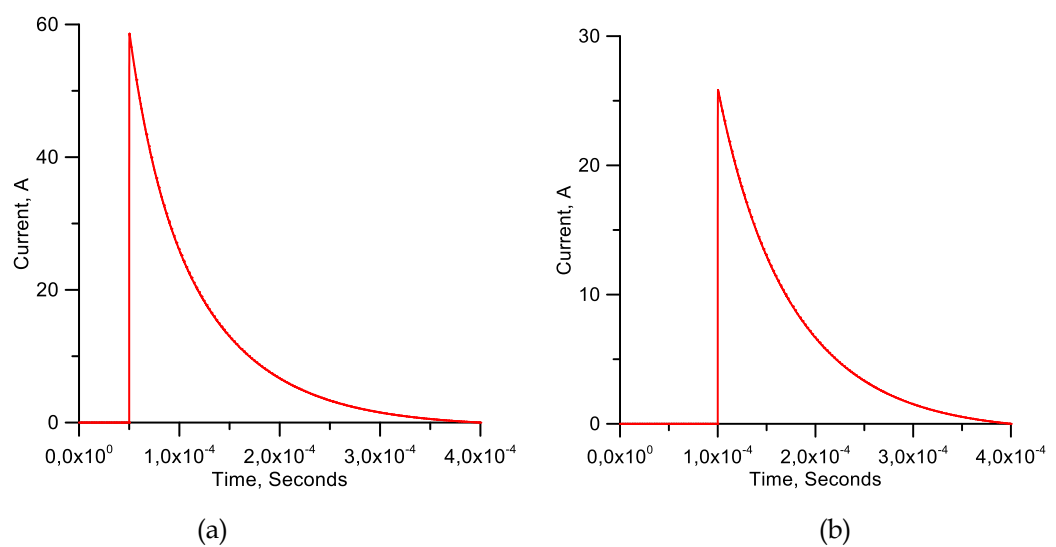

**Figure 4.** The mini return stroke current at a height of (a) 5 km and (b) 10 km.

An analysis similar to that conducted in the previous paragraph can be done to estimate how the current generated by the induction process as the macro channel travels towards ground is distributed along the channel. The procedure is identical to the one presented earlier except that the discharge current is replaced by a charging current and this charging current travels upwards instead of downwards as in the case of the neutralization process [10]. The average currents propagating along the macro channel at three heights are shown in Figure 5. The following points are worth noting when interpreting the data shown in Figure 5. Consider a given height from ground level. The current at that height starts to flow immediately after the macro has passed that height. Since charge continues to accumulate at sections of the macro channel below this height, the current at that height will continue to flow until the current generated by the lowest point of the macro channel reaches that height. In the diagram, the current at each height is shown only until the arrival of the current generated by the lowest point of the macro channel. Immediately after this time, the mini return stroke will pass through that height. It is important also to point out that the current at different heights starts at different times (i.e., at the time when the macro passes that height). However, for clarity, these delays in current starting times are removed in Figure 5. Earlier we have referred to the current depicted in Figure 5 as the average current generated by the charging process. The reason for this definition is the following. In estimating the current at any given height, we have used the charge density depicted in Figure 2. Actually, the charge densities presented in Figure 2 are the net charge deposited at any given level. However, as the macro channel passes a given height, more charge is deposited at that level initially and as the channel extends forward, the charge is gradually reduced reaching the values depicted in Figure 2. Thus, the charging current is actually bipolar, initially depositing more charge and subsequently removing some of the charge as the channel extends forward. For this reason, what is represented in Figure 5 is the average charging current.

Note that the current propagating along the dark matter channel is in the order of several tens of amps. These currents could help in maintaining the conductivity of the macro channel.

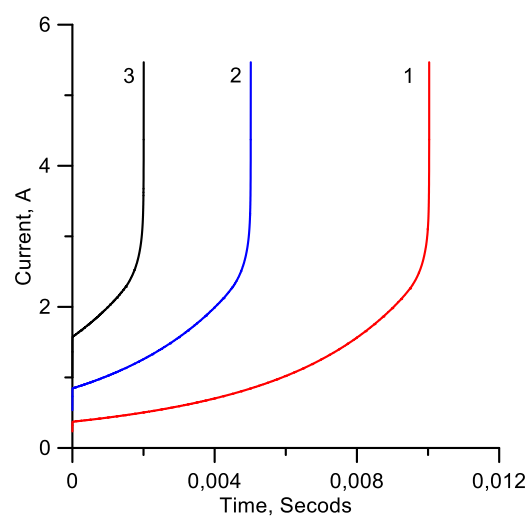

**Figure 5.** The current induced in the macro channel after the creation of the macro channel at different heights. The speed of propagation of the macro was assumed to be $10^6$ m/s. (1) 10 km; (2) 5 km (3) 2 km. For clarity, the time delays involved in the initiation of current at different levels were removed.

## 5. Discussion

Initiation of lightning flashes by energetic particles or cosmic radiation is a hot topic at present. Gurevich [11] presented a theory that detailed the creation of compact cloud discharges or narrow bipolar events by relativistic runaway electrons created by cosmic radiation. According to this theory, an Extensive Atmospheric Shower (EAS) consisting of Relativistic Runaway Electron Avalanches (RREA) triggered by a cosmic ray increases the ion and electron densities in a cloud, triggering electrical events that give rise to compact cloud discharges, whose radiation fields are similar to or more energetic than the radiation fields of return strokes but with a shorter duration. The possibility of such events leading to the initiation of normal lightning flashes is still under discussion within the scientific community. The research work carried out by Nicole et al. [12] demonstrates that RREA give rise to gamma ray glows near the top of thunderclouds. According to the above reference, these glows are capable of discharging the upper positive charge layer in clouds and they have to be included in the mechanisms involved in the charging and discharging of the thundercloud. However, whether RREA and energetic gamma rays produced inside the thundercloud or cosmic rays that originate outside the Earth's atmosphere could give rise to an ionization path all the way from the cloud to ground paving the way for a lightning flash has not yet been investigated in the literature. If they can give rise to conducting channels similar to those assumed to be created by the passage of macros through the atmosphere, they might also give rise to mini-lightning flashes. Simultaneous measurements of electromagnetic radiation and X-rays or gamma radiation from such paths would be used to distinguish whether the ionization path is created by a relativistic ionization event or by a macro.

The mini-lightning flash produced by a dark matter particle will have a peak current of about 200 A and it has duration of about 400 μs. The total charge transported is about .01 C. Now, the magnitude of continuing currents in lightning flashes can be in the range of 10 to 250 A. The experimental observations show that channels carrying continuing currents as low as 10 A could be optically recorded [13]. Moreover, the currents in laboratory sparks are also located in this range and they generate visible optical radiation and the sound created by them can be heard several hundred meters away. The mini-lightning channel being much longer, it may generate an audible thunder-signature. The channel will be visible during daytime only in its vicinity and, during nighttime, over a longer distance. As in the case of macro-generated lightning striking through a thundercloud, the channel will be completely straight. Of course, in the case of a macro channel associated with a thundercloud, there is a possibility for the leader discharge to shoot out from the tip of the macro channel, thus camouflaging the channel geometry. In the case of mini-lightning flashes generated by a macro, the induced charge is not large enough to generate an electrical discharge shooting out through the macro channel. Thus, the channel will be completely straight. Moreover, channel geometry will not be affected by connecting leaders because the induced charge would not be large enough to generate electric fields at ground level capable of launching connecting leaders. Thus, if macro channels exist in the atmosphere, one should be able to identify them from the mini-lightning flashes they generate when they are passing through fair weather regions without the aid of thunderclouds.

In the analysis, we have assumed that the macro channel remains conducting as it propagates from the ionosphere to ground. From Figure 1b, one can see that the contribution to the atmospheric potential is coming from the atmosphere, which is located below about 20 km. Thus, the results presented here would be valid even if the macro channel remains conducting only over several km of its travel to ground where the atmospheric electric field is strongest. Now, the question is whether the channel created by macros will remain conducting over several kilometers from ground level. The problem we have is the following. The passage of the macro will give rise to a conducting channel of a radius of a cm or so at a certain temperature. Very soon after its creation, the channel starts to decay. If this decay rate is slow and the speed of the macro is fast enough, then the length of the conducting channel remains at a few kilometers, which could be polarized by the external field. However, if the decay rate is fast and the speed of the macro is small, only a small portion of the channel remains conducting behind the moving macro. This inhibits or limit the amount of polarization charges deposited along the channel. Unfortunately, Starkman et al. [1] did not provide a plot of the temperature profile of the macro channel as a function of time. However, after taking into account the possible losses in the channel, these authors estimate that the linear electron density of the macro channel is about 6 x $10^{13}$/cm. In atmospheric air, this electron density corresponds to about 4000 K. At this temperature, the conductivity of the plasma is about 2.4 S/m. How fast the macro channel will be polarized by the external electric field depends on how fast the temperature and hence the conductivity of the macro channel decreases with time. Uman and Voshall [14] presented calculations to illustrate the decay of lightning channels at different temperatures with time. According to their results, a 1-cm diameter channel will decay from about 4000 K to 2000 K in about 50 ms and a 2 cm channel decays from 4000 K to about 3000 K in about 50 ms. Calculations of Uman and Voshall [14] also show that a 10 cm thick channel will decay at a much slower rate. Thus, it is reasonable to assume that the conductivity of the macro channel remains close to about 2.4 S/m for tens of milliseconds. As mentioned previously, we have assumed that the radius of the macro channel is about 10 cm. According to Starkman et al. [1] a macro moving at a speed of about $2.5 \times 10^5$ m/s will give rise to a conducting channel of about 2 cm thick. The radius of the channel increases with the square of the speed of the macro and in order to generate a 10 cm thick channel the macro has to travel at a speed of about $5 \times 10^5$ m/s. According to the speed distribution of macros given by Starkman et al. [1], about 10% of the number of macros propagating at speeds larger than $2.5 \times 10^5$ m/s will be propagating at speeds larger than about $5 \times 10^5$ m/s. As the speed of the macro decreases the thickness of the channel decreases and so does the charge induced on the channel. Moreover, as we will see later the channel resistance will also increase with decreasing radius thus further inhibiting the charging process. In tens of milliseconds, a macro moving at $5 \times 10^5$ m/s will travel a distance of several kilometers and, therefore, one can expect that at least a few km of the channel remain at a conductivity of about 2.4 S/m. The faster the macro, the longer will be the channel section available for polarization. With this conductivity, a channel of 10 cm radius will have a channel resistance of about 10 Ohms/m. Along such a channel, a 100 V/m electric field can induce a current of about 10 A. The distance travelled by a macro in 10 microseconds is about five meters and this induced current is large enough to charge the channel to the charge densities estimated earlier. Of course, as mentioned earlier, as the speed of the macro decreases, the channel radius decreases while the channel resistance increases and so does the induced currents in the macro channel.

In the calculations presented by Uman and Voshall [14], there is no energy input into the channel while it decays. However, in our case, there is some energy input into the channel by the induced currents during its decaying stage. The magnitude of the induced currents we have estimated here (see Figure 5) is in the order of a few Amperes. The important point here is that a conducting channel of radius of about a 10 cm or more is available for this induced currents to flow. This is very different to a channel in air created by a discharge having a current of about a few A. In such a discharge, the channel radius available for the current flow is significantly smaller and the channel resistance would be significantly higher [15]. However, in our case, the resistance of the channel is about 10 Ohm/m (due to its larger radius) and the energy dissipated by a 5 A induced current (see Figure 5) flowing along this channel for about 10 ms is about 2.5 J. This energy input into the channel by the induced currents will retard the decay of the channel temperature and conductivity. This further strengthens the assumption of having a several kilometers long conducting channel during the passage of the macro.

In order to create a macro lightning in association with thunderclouds, the path of the macro channel should cross the thunderstorm region. Assuming that the size of a thunderstorm is about 20 km in diameter and at any given time there are about 2000 active thunderstorms, Starkman et al. [1] estimated that the area covered by thunderstorms is about 0.003 of the total surface area of the Earth. Using this fraction, Starkman et al. [1] have estimated that the frequency of macro-generated lightning is about $10^{-6}$ /s. On the other hand, mini-lightning flashes produced by macros in fair weather have almost all the surface area of the Earth to manifest themselves. Taking also into account that we are considering macros moving with speeds faster than about $5 \times 10^5$ m/s, the frequency of macro produced mini-lightning would be about $3 \times 10^{-5}$/s, which is about 30 times larger than the thundercloud mediated macro channels as estimated by Starkman et al. [1]. This high frequency makes it more favorable to look for mini-lightning flashes caused by macros than the lightning flashes produced by macros in association with thunderclouds. Moreover, the latter could be camouflaged as normal lightning due to various factors discussed earlier. Furthermore, they could also be hidden within the flux of normal lightning flashes which is about $10^8$ times more frequent.

Another physical event that gives rise to an ionized channel in the atmosphere is the passage of meteorites across the Earth's atmosphere. Since these trails are conducting, they lead to the reflection of electromagnetic fields and hence could be identified by radar.

Recently, it has been suggested by Kelley and Price [16] that meteorite trails not only reflect electromagnetic radiation, but they also generate electromagnetic waves. That study proposes the interaction of the free electrons in the channel with the Earth's magnetic field as a possible source of these electromagnetic waves. It is also possible that the trail of the meteorites reaching all the way to ground could be polarized by the external electric fields and generate electromagnetic radiation. However, since the speed of these meteorites is on the order of 3 x10$^4$ m/s or larger (the speed of the meteorite close to the ground could be smaller than this value due to attenuation caused by air resistance), which is smaller than the ones assumed for the macros, the decay of the conductivity with time would make only a small portion of the channel from the tip of the meteor be conducting, and only this portion could be polarized by external fields. In this respect, the meteors reaching the ground can be treated more as moving dipoles where the charge of the dipole is confined to the region of the channel in the vicinity of the head of the meteor. It is therefore doubtful that a meteor striking the ground can also give rise to a mini lightning flash.

As discussed above, we have estimated that macros could generate about 3x10$^{-4}$ mini-lightning flashes per second. Interestingly, this number is similar to the number of meteorites that are estimated to travel all the way to ground [17]. It is important to note that the surface area of the Earth that is populated by humans (i.e., the urban areas) is less than about 1% and, therefore, most of these mini lightning flashes, if they occur, will pass through the atmosphere unnoticed. Even if it happens in an urban region, the chances are high that the event will be classified as a small meteor striking the earth. This is the case because a small meteor travelling at a speed faster than the speed of sound will generate an explosive sound and the vaporization of the material will give rise to an illuminated channel. Whether the process is created by a macro or a meteorite can be decided only by appealing to the electrical measurements.

## 6. Conclusions

The results presented in this paper show that macroscopic dark matter channels created in fair weather regions of the Earth could give rise to mini-lightning flashes with peak currents in the order of 100 – 300 A. Judging from the experimental observations available on electrical discharges carrying similar currents in the atmosphere, these mini lightning flashes should be visible to the naked eye and they should create a thunder signature which is audible for observers located in its vicinity. Moreover, the electric and magnetic field signatures of these mini-lightning flashes would be similar to those of normal lightning return strokes but with a much smaller amplitude. The current lightning location system technology could therefore be suitably adapted to detect and locate these mini-lightning flashes.